\documentclass{ws-ijmpe}
\usepackage[super,compress]{cite}

\newcommand{\nn}{\nonumber}

\newcommand{\beq}{\begin{equation}}
\newcommand{\eeq}{\end{equation}}
\newcommand{\beqa}{\begin{eqnarray}}
\newcommand{\eeqa}{\end{eqnarray}}
\newcommand{\bd}[1]{ \mbox{\boldmath $#1$}}

\begin{document}
\def\ii{\'\i}

\markboth{O. Civitarese, P. O. Hess, D. A. Amor-Quiroz}
{Generalized variational procedure: An application to non-perturbative QCD}

\title{Generalized variational procedure: An application to non-perturbative QCD}

\author{\footnotesize O. CIVITARESE}

\address{
Departamento de F\ii sica, Universidad Nacional de La Plata,
C.C.67 (1900), La Plata, Argentina\\
osvaldo.civitarese@fisica.unlp.edu.ar
}

\author{\footnotesize P. O. HESS}

\address{
Instituto de Ciencias Nucleares, UNAM, Circuito Exterior, C.U.,
A.P. 70-543, 04510 M\'exico, D.F., Mexico \\
and \\
Frankfurt Institute for Advanced Studies, Johann Wolfgang Goethe Universit\"at,
Ruth-Moufang-Str. 1, 60438 Frankfurt am Main, Germany \\
and \\
GSI Helmholtzzentrum f\"uer Schwerionenforschung GmbH,
Max-Planck-Str. 1, 64291 Darmstadt, Germany\\
hess@nucleares.unam.mx
}

\author{\footnotesize D. A. AMOR-QUIROZ}

\address{
Instituto de Ciencias Nucleares, UNAM, Circuito Exterior, C.U.,
A.P. 70-543, 04510 M\'exico, D.F., Mexico \\
arturo.amor@nucleares.unam.mx
}

\maketitle

\begin{history}
\received{(received date)}
\revised{(revised date)}
%\accepted{(Day Month Year)}
%\comby{(xxxxxxxxxx)}
\end{history}

\begin{abstract}
We present a generalized variational procedure oriented to the
algebraic solution of many body Hamiltonians expressed in bosonic
and fermionic variables. The method specializes in the
non-perturbative regime of the solutions. As an example, we focus on
the application of the method to non-perturbative QCD.
%\pacs{10.,20.,21.60.De}
\end{abstract}


\ccode{PACS numbers: 10.,20.,21.60.De}

\vskip 1cm

\section{Introduction}
Variational procedures have proved to be useful tools in the
treatment of various quantum many-body problems, from molecular to
hadron physics \cite{ring,greiner-2}. As an example, we shall
mention the case of the low-energy hadronic spectrum, as described
by QCD \cite{td-lee} in its non-perturbative regime, where one has
to deal, simultaneously, with confined fermions (quarks and
anti-quarks) and bosons (gluons). To find there an effective method
to diagonalize the QCD Hamiltonian is not at all trivial. One often
recurs to trial states, as the coherent state defined in
\cite{stump,kogan,adam-trial-1,adam-trial-2}, which give some
insight on the vacuum structure of QCD. In \cite{paper-2010} a first
step in this direction was reported. The Coulomb interaction was
approximated by a contact interaction and a semi-analytic solution
of the QCD Hamiltonian was found. Based on these results, we aim at
developing  a variational method to diagonalize the complete QCD
Hamiltonian \cite{adam-swanson} within the basis constructed in
\cite{paper-2010}.

The paper is organized as follows: In section II a boson system is
considered. The transformation to new quasi-particles is presented
and the nature of the new vacuum state is discussed. In section III
the same is done for a fermionic many-body system. In section IV we
show the relation between the present method and the coherent state
used in a variational treatment of QCD
\cite{adam-trial-1,adam-trial-2}. Finally, in section V the
conclusions are drawn.

\section{A system of many bosons}
Let us define the creation and annihilation operators, ${\bd
b}^\dagger_{\alpha a}$ and ${\bd b}^{\alpha a}$, where the two
indexes are to be associated with different degrees of freedom. In
the case of QCD, the index $a$ may refer to color and $\alpha$ to
flavor, spin and orbital degrees of freedom. Lower and upper indices
denote the difference under transformations of the creation and
annihilation operators. These operators obey the commutation relation

\beqa 
\left[ {\bd b}^{\alpha a} , {\bd b}^\dagger_{\beta b} \right]
& = & \delta_{\alpha\beta}\delta_{ab} ~~~. 
\label{com-bos} 
\eeqa 
We introduce the following general ansatz for the transformation to new
creation and annihilation operators
\beqa {\bd P}^\dagger _{\alpha ,
a} & = & \frac{1}{\sqrt{2}}\sum_\beta \left( M_{\alpha}^{~\beta}
{\bd b}^\dagger_{\beta a} + N_{\alpha}^{~\beta} {\bd b}_{\beta a}
\right)
\nonumber \\
{\bd P}^{\alpha , a} & = & \frac{1}{\sqrt{2}}\sum_\beta \left(
M^{~\alpha}_{\beta} {\bd b}^{\beta a} - N^{~\alpha}_{\beta} {\bd
b}^{\dagger\beta a}  \right) ~~~. \label{eq1} \eeqa The indices
$\alpha$ and $\beta$ refer to the spatial quantum numbers, {\it
except the magnetic projection of total spin}, and $a$ is a short
hand notation for the sum of the color index and the magnetic
projection of the spin. By construction the matrices $M$ and $N$
transform under raising and lowering of the indices in the same way
as the creation and annihilation operators, to preserve the
transformation properties of the ${\bd P}$ operators. In QCD the
indices $\alpha$ and $\beta$ run over the values which are allowed
by the Coulomb condition (i.e: restricted to transversal
space-components) and they can be divided into several other
indexes, as done in \cite{paper-2010,tochtli-th}. For example, \beqa
\alpha & \rightarrow & \Xi (N,L,1)J ~~~, \label{1aa} \eeqa where
$\Xi$ denotes either the ${\cal E}$ (electric) or ${\cal M}$
(magnetic) modes, $N$ is the principal quantum number (when the
harmonic oscillator is used as a basis), $L$ the orbital angular
momentum and $J$ the total spin. For the boson operators we use the
phase convention \beqa {\bd b}^{\alpha a} & = & (-1)^{\varphi
(\alpha )+\varphi (a)} {\bd b}_{{\bar \alpha}{\bar a}} \label{eq3}
\eeqa under raising and lowering indices, where the bar over an
index denotes its conjugate component \footnote{In this notation,
the angular momentum projection $M$ goes into $-M$ and in $SU(3)$
the color-hypercharge and color-isospin  $(Y,T,T_z)$ goes into
$\rightarrow$ $(-Y,T,-T_z)$.}\cite{draayer-aki,draayer-bahri,jutta}.
With the phase transformation (\ref{eq3}), we have \beqa
M^{~\alpha}_\beta & = & (-1)^{\varphi (\alpha) + \varphi (\beta) }
M_{~~{\bar \alpha}}^{{\bar \beta}}
\nonumber \\
N^{~\alpha}_\beta & = & (-1)^{\varphi (\alpha) + \varphi (\beta) }
N_{~~{\bar \alpha}}^{{\bar \beta}} ~~~. \label{eq3a} \eeqa Next, we
investigate the properties of the ${\bd P}$ operators under
commutation. They are boson operators, which gives conditions to the
$M$ and $N$ matrices. Then,
\beqa &
\delta_{\alpha\beta}\delta_{ab}~=~ \left[ {\bd P}^{\alpha a},{\bd
P}^\dagger_{\beta b} \right] &
\nonumber \\
& = &
\nonumber \\
& \frac{1}{2}
\sum_{\gamma_1\gamma_2}
M^{~\alpha}_{\gamma_1} M_\beta^{~\gamma_2}
\left[ {\bd b}^{\gamma_1 a},{\bd b}^\dagger_{\gamma_2 b} \right]
&
\nonumber \\
&
+ \frac{1}{2}\sum_{\gamma_1\gamma_2}
M^{~\alpha}_{\gamma_1} N_\beta^{~\gamma_2}
\left[ {\bd b}^{\gamma_1 a},{\bd b}_{\gamma_2 b} \right]
&
\nonumber \\
&
- \frac{1}{2}\sum_{\gamma_1\gamma_2}
N^{~\alpha}_{\gamma_1} M_\beta^{~\gamma_2}
\left[ {\bd b}^{\dagger\gamma_1 a},{\bd b}^\dagger_{\gamma_2 b} \right]
&
\nonumber \\
&
- \frac{1}{2}\sum_{\gamma_1\gamma_2}
N^{~\alpha}_{\gamma_1} N_\beta^{~\gamma_2}
\left[ {\bd b}^{\dagger\gamma_1 a},{\bd b}_{\gamma_2 b} \right]
&
\nonumber \\
& = &
\nonumber \\
& \delta_{ab}\sum_{\gamma} \frac{1}{2} \left( M^{~\alpha}_{\gamma}
M_\beta^{~\gamma} + N^{~\alpha}_{\gamma} N_\beta^{~\gamma} \right) &
~~~. \label{eq3b} \eeqa which implies
\beqa \sum_{\gamma}
M_\beta^{~\gamma} M^{~\alpha}_{\gamma} & = & \delta_{\alpha\beta}
\nonumber \\
\sum_{\gamma}
N_\beta^{~\gamma} N^{~\alpha}_{\gamma} & = & \delta_{\alpha\beta}
~~.
\label{eq3c}
\eeqa
One is free to choose $N_\alpha^{~\beta}$ = $M_\alpha^{~\beta}$, which we adopt from
here on.

\subsection{The trial state for the bosonic many-body system}

For the trial state we define a new vacuum $\mid {\tilde 0}
\rangle$, which satisfies \beqa {\bd P}^{\alpha a} \mid {\tilde 0}
\rangle & = & 0 ~~~. \label{eq20} \eeqa In order to calculate matrix
elements of the boson creation and annihilation operators, Eq.
(\ref{eq1}) has to be inverted, giving
\beqa \left( {\bd
b}^\dagger_{\beta a} + {\bd b}_{\beta a}  \right) & = &
\sqrt{2}\sum_{\alpha} M_\beta^{~\alpha} {\bd P}^\dagger_{\alpha b}
\nonumber \\
\left( {\bd b}_{\beta a} - {\bd b}^\dagger_{\beta a}  \right) & = &
\sqrt{2}\sum_{\alpha} M^\alpha_{~\beta} {\bd P}_{\alpha a} ~~~.
\label{eq21} \eeqa and, from these equations one gets
\beqa {\bd
b}^\dagger_{\beta b} & = & \frac{1}{\sqrt{2}} \sum_\alpha \left(
M_\beta^{~\alpha}{\bd P}^\dagger_{\alpha b} - M_{~\beta}^\alpha{\bd
P}_{\alpha b} \right)
\nonumber \\
{\bd b}_{\beta b} & = & \frac{1}{\sqrt{2}} \sum_\alpha \left(
M_\beta^{~\alpha}{\bd P}^\dagger_{\alpha b} + M_{~\beta}^\alpha{\bd
P}_{\alpha b} \right) \label{eq23} \eeqa
The above transformation is
more general. Under scaling, and preserving the commutation
relations, we write \beqa {\bd b}^\dagger_{\alpha a} & \rightarrow &
\sqrt{\lambda_\alpha} {\bd b}^\dagger_{\alpha a}
\nonumber \\
{\bd b}^{\alpha a} & \rightarrow &
\frac{1}{\sqrt{\lambda_\alpha}} {\bd b}^{\alpha a}
~~~,
\label{eq27}
\eeqa

The scaling $\lambda_\alpha$ may depend on the index $a$, a choice
leading to a trial state which does not have definite color or spin.
A further generalization of the transformation is possible, to
include a displacement of the operators
\beqa {\bd
b}^\dagger_{\alpha a} & \rightarrow & \sqrt{\lambda_\alpha} {\bd
b}^\dagger_{\alpha a} - \eta_{\alpha a}
\nonumber \\
{\bd b}^{\alpha a} & \rightarrow & \frac{1}{\sqrt{\lambda_\alpha}}
{\bd b}^{\alpha a} - \eta^{\alpha a} ~~~. \label{eq32b} \eeqa The
$\eta_{\alpha a}$ are tensors and transform in the same way as the
${\bd b}$-operators.
\subsection{Nature of the boson vacuum state}

With the redefinition (\ref{eq27}), Eq. (\ref{eq21}) changes to
\beqa \left( \sqrt{\lambda_\alpha} {\bd b}^\dagger_{\alpha a} -
\frac{1}{\sqrt{\lambda_\alpha}}{\bd b}_{\alpha a}  \right) & = &
-\sqrt{2}\sum_{\beta} M^{\beta}_{~\alpha} {\bd P}_{\beta a} ~~~.
\label{eq28} \eeqa Applying it to the new vacuum state $\mid {\tilde
0}\rangle$, gives \beqa \left( \sqrt{\lambda_\alpha} {\bd
b}^\dagger_{\alpha a} - \frac{1}{\sqrt{\lambda_\alpha}}{\bd
b}_{\alpha a}  \right) \mid {\tilde 0} \rangle & = & 0 ~~~.
\label{eq29} \eeqa or \beqa {\bd b}_{\alpha a} \mid {\tilde 0}
\rangle  & = & \lambda_\alpha {\bd b}^\dagger_{\alpha a} \mid
{\tilde 0} \rangle ~~~. \label{eq30} \eeqa The annihilation
operators act as derivatives to the creation operators, i.e., \beqa
{\bd b}_{\alpha a} & \rightarrow & \frac{\partial} {\partial {\bd
b}^{\dagger\alpha a}} ~~~. \label{eq30a} \eeqa This provides, for
the component $(\alpha~a)$, the expression \beqa \mid {\tilde 0}
\rangle & \sim & e^{\frac{1}{2}\lambda_\alpha {\bd
b}^\dagger_{\alpha a} {\bd b}^{\dagger\alpha a}} \mid 0 \rangle ~~~.
\label{eq31} \eeqa and it leads to the solution
\beqa
\mid {\tilde
0} \rangle & \sim & e^{\frac{1}{2}\sum_\alpha\lambda_\alpha \left(
{\bd b}^\dagger_{\alpha} \cdot {\bd b}^{\dagger\alpha} \right)} \mid
0 \rangle ~~~,
\label{eq32}
\eeqa
where $\left( {\bd b}^\dagger_{\alpha} \cdot {\bd b}^{\dagger\alpha} \right)$ = $
\sum_a {\bd b}^\dagger_{\alpha a} {\bd b}^{\alpha a}$.
The introduction
of the parameters $\lambda_\alpha$ gives us a further freedom in the
variational procedure.  The ${\bd P}$-operators can be cast into the
form

\beqa
{\bd P}^\dagger_{\alpha a} & = & \frac{1}{\sqrt{2}}
\sum_\beta M_\alpha^{~\beta} \left( \sqrt{\lambda_\beta}{\bd
b}^\dagger_{\beta a} + \frac{1}{\sqrt{\lambda_\beta}}{\bd b}_{\beta
a} \right)
\nonumber \\
{\bd P}_{\alpha a} & = & \frac{1}{\sqrt{2}}\sum_\beta
M_{~~\alpha}^\beta \left( \sqrt{\lambda_\beta}{\bd b}_{\beta a} -
\frac{1}{\sqrt{\lambda_\beta}}{\bd b}^\dagger_{\beta a} \right) ~~~.
\label{eq32aa}
\eeqa
This allows to vary $M_\alpha^{~\beta}$ (and it
complex conjugate $M^{\alpha}_{~\beta}$), and $\lambda_\alpha$.
Using the more general definition of (\ref{eq32b}) does not change
equations (\ref{eq28}) and (\ref{eq29}). Thus, the structure of the
trial state will stay the same, because the differential equation in
terms of the ${\bd b}$-operators will be the same.  Eq. (\ref{eq23})
will change to

\beqa 
{\bd b}^\dagger_{\beta b} & = &
\frac{1}{\sqrt{2\lambda_\beta}} \sum_\alpha \left(
M_{\beta}^{~\alpha}{\bd P}^\dagger_{\alpha b} -
M_{~~\beta}^\alpha{\bd P}_{\alpha b} \right) + \frac{\eta_{\beta
b}}{\sqrt{\lambda_\beta}}
\nonumber \\
{\bd b}_{\beta b} & = & \sqrt{\frac{\lambda_\beta}{2}}
\sum_\alpha
\left( M_\beta^{~\alpha}{\bd P}^\dagger_{\alpha b}
+ M_{~~\beta}^\alpha{\bd P}_{\alpha b} \right)
+ \sqrt{\lambda_\beta}\eta_{\beta b}
\nonumber \\
\label{eq32c} 
\eeqa 
and the equations for the ${\bd P}$-operators
change to

\beqa 
{\bd P}^\dagger_{\alpha a} & = &
\frac{1}{\sqrt{2}}\sum_\beta M_\alpha^{~\beta} \left(
\sqrt{\lambda_\beta}{\bd b}^\dagger_{\beta a} +
\frac{1}{\sqrt{\lambda_\beta}}{\bd b}_{\beta a} \right)
\nonumber \\
&&
+\sqrt{2}
\sum_\beta M_{\alpha}^{~\beta}
\eta_{\beta a}
\nonumber \\
{\bd P}_{\alpha a} & = & \frac{1}{\sqrt{2}}\sum_\beta
M_{~~\alpha}^\beta \left( \sqrt{\lambda_\beta}{\bd b}_{\beta a} -
\frac{1}{\sqrt{\lambda_\beta}}{\bd b}^\dagger_{\beta a} \right) ~~~.
\label{eq32aaa} \eeqa For the annihilation operator no contribution
of $\eta_{\alpha a}$ enters its definition. The parameters to vary
are: \beqa & M_\alpha^{~\beta} ~,~ \lambda_\alpha ~{\rm and}~
\eta_{\alpha a} ~~~. 
\label{eq32d} 
\eeqa 
and the use of
$\eta_{\alpha a}$ allows us to shift the operators by their vacuum
expectation values, as done in presence of spontaneously broken
symmetries.

\section{A system of many fermions}
In this section, we extend the above consideration to fermions. We
use the notation \beqa {\bd b}^\dagger_{\alpha (1,0) a, j m} ~,~
{\bd d}^\dagger_{\alpha (0,1) {\bar a}, jm} ~~~. \label{32e} \eeqa
The ${\bd b}$-operators refer to quarks and the ${\bd d}$-operators to
anti-quarks. They transform differently under $SU(3)$. While the
quarks transform with respect to color as an (1,0) irreducible
representation (irrep), and the anti-quarks as the irrep-(0,1). The
same transformation properties apply for the annihilation operators
\beqa {\bd b}^{\alpha (1,0) a, j m} ~,~ {\bd d}^{\alpha (0,1) {\bar
a}, jm} ~~~. \label{32f} \eeqa The indices $\alpha$, $\beta$ refer
to numbers like the principal quantum number and other orbital
indices, except color and spin. They can be treated as cartesian.
Thus, the phase factors $\varphi (\alpha )$ of the previous sections
are set to zero, i.e. $(-1)^{\varphi (\alpha )}=+1$. In the Appendix
we derive the properties of the creation and annihilation operators
under raising and lowering indices, using the phase convention of
\cite{draayer-aki,draayer-bahri}. Here, we resume the results: \beqa
{\bd b}^{\dagger \alpha (0,1) {\bar a}, jm} & = & (-1)^{\chi_a +j+m}
{\bd b}^\dagger_{\alpha (1,0) a, j -m}
\nonumber \\
{\bd b}^{\dagger}_{\alpha (1,0) a, jm} & = &
(-1)^{\chi_a +j-m} {\bd b}^{\dagger \alpha (0,1) {\bar a}, j -m}
\nonumber \\
{\bd b}^{\alpha (1,0) a, jm} & = &
(-1)^{\chi_a +j-m} {\bd b}_{\alpha (0,1) {\bar a}, j -m}
\nonumber \\
{\bd b}_{\alpha (0,1) {\bar a}, jm} & = &
(-1)^{\chi_a +j+m} {\bd b}^{\alpha (1,0) a, j -m}
~~~,
\nonumber \\
\label{32h} \eeqa with $\chi_a = \frac{\lambda
-\mu}{3}+\frac{Y}{2}+T_z$, $Y$ is the hypercharge and $T_z$ the
third component of the isospin. A similar property holds for the
anti-particle operators, i.e., \beqa {\bd d}^{\alpha (0,1) {\bar a},
jm} & = & (-1)^{\chi_a +j-m} {\bd d}_{\alpha (1,0) a, j -m}
\nonumber \\
{\bd d}_{\alpha (1,0) a, jm} & = &
(-1)^{\chi_a +j+m} {\bd d}^{\alpha (0,1) {\bar a}, j -m}
\nonumber \\
{\bd d}^{\dagger \alpha (1,0) a, jm} & = &
(-1)^{\chi_a +j+m} {\bd d}^{\dagger}_{\alpha (0,1) {\bar a}, j -m}
\nonumber \\
{\bd d}^{\dagger}_{\alpha (0,1) {\bar a}, jm} & = &
(-1)^{\chi_a +j-m} {\bd d}^{\dagger \alpha (1,0) a, j -m}
~~~.
\nonumber \\
\label{32i} \eeqa
and \beqa \left\{ {\bd b}^{\alpha (1,0) a, j_1m_1}
, {\bd b}^\dagger_{\beta (1,0) b, j_2m_2} \right\} & = &
\delta_{ab}\delta_{\alpha\beta} \delta_{j_1j_2} \delta_{m_1m_2}
\nonumber \\
\left\{ {\bd d}^{\alpha (0,1) {\bar a}, j_1m_1} , {\bd
d}^\dagger_{\beta (0,1) {\bar b}, j_2m_2} \right\} & = &
\delta_{ab}\delta_{\alpha\beta} \delta_{j_1j_2} \delta_{m_1m_2} ~~~.
\label{eq33} \eeqa (all other anti-commutators vanish). To perform
the mapping onto new fermion operators, we follow the same procedure
as for (\ref{eq1}). These new fermion operators ${\bd P}^\dagger$
and ${\bd P}$ are given by  \footnote{For simplicity, we keep the
same notation of the previous section, and the fermionic or bosonic
character of the operators will be specified when needed.} \beqa
\bd{P}^{\dagger}_{\alpha (1,0) a,jm} &=& \frac{1}{\sqrt{2}}
\sum_{\beta} M^{~\beta}_{\alpha} \left( \bd{b}^{\dagger}_{\beta
(1,0) a,jm} +  \bd{d}_{\beta (1,0) a,jm} \right)
\nonumber \\
\bd{P}^{\alpha (1,0) a,jm} &=& \frac{1}{\sqrt{2}}
\sum_{\beta} M^{~\alpha}_{\beta}
\left( \bd{b}^{\beta (1,0) a,jm}
+  \bd{d}^{\dagger \beta (1,0) a,jm} \right)
\nonumber \\
\bd{D}^{\dagger}_{\alpha (0,1) {\bar a},jm} &=& \frac{1}{\sqrt{2}}
\sum_{\beta} M^{\beta}_{~\alpha}
\left( \bd{d}^{\dagger}_{\beta (0,1) {\bar a},jm}
+  \bd{b}_{\beta (0,1) {\bar a},jm} \right)
\nonumber \\
\bd{D}^{\alpha (0,1) {\bar a},jm} &=& \frac{1}{\sqrt{2}}
\sum_{\beta} M^{\alpha}_{~\beta}
\left( \bd{d}^{\beta (0,1) {\bar a},jm}
+  \bd{b}^{\dagger \beta (0,1) {\bar a},jm} \right)
~~~,
\nonumber \\
\label{eq34} \eeqa Lowering the indices of the annihilation
operators gives (see the Appendix and (\ref{32h})). \beqa
\bd{P}_{\alpha (0,1) {\bar a},jm} &=& \frac{1}{\sqrt{2}}
\sum_{\beta} M^{\beta}_{~\alpha} \left( \bd{b}_{\beta (0,1) {\bar
a},jm} -  \bd{d}^{\dagger}_{\beta (0,1) {\bar a},jm} \right)
\nonumber \\
\bd{D}_{\alpha (1,0) a,jm} &=& \frac{1}{\sqrt{2}}
\sum_{\beta} M^{~~\beta}_{\alpha}
\left( \bd{d}_{\beta (1,0) a,jm}
-  \bd{b}^{\dagger}_{\beta (1,0) a,jm} \right)
~~~.
\nonumber \\
\label{eq34a} \eeqa One reason for this ansatz is that there is no
mixing of color (both the quark creation-operator and the anti-quark
annihilation operator with lower indexes belong to the same irrep).
Note that these definitions preserve anti-commutation relations.

\subsection{The trial state for the fermionic many-body system}
With respect to the new vacuum, we require that \beqa \bd{P}_{\alpha
(0,1) {\bar a}, jm}  | \tilde{0} \rangle & = & 0
\nonumber \\
\bd{D}_{\alpha (1,0) a, jm}  | \tilde{0} \rangle & = & 0 ~~~.
\label{eq35} \eeqa
By inversion we get \beqa \bd{b}^{\dagger}_{\beta
(1,0) a,jm} + \bd{d}_{\beta (1,0) a,jm}  &=& \sqrt{2} \sum_{\alpha}
M^{~\alpha}_{\beta}  \bd{P}^{\dagger}_{\alpha (1,0) a,jm}
\nonumber \\
\bd{b}_{\beta (0,1) {\bar a},jm} - \bd{d}^\dagger_{\beta (0,1) {\bar a},jm}
&=&
\sqrt{2}
\sum_{\alpha} M^{\alpha}_{~\beta}  \bd{P}_{\alpha (0,1) {\bar a},jm}
\nonumber \\
\bd{d}^{\dagger}_{\beta (0,1) {\bar a},jm} + \bd{b}_{\beta (0,1) {\bar a},jm}
&=& \sqrt{2}
\sum_{\alpha} M^{\alpha}_{~\beta}  \bd{D}^{\dagger}_{\alpha (0,1) {\bar a},jm}
\nonumber \\
\bd{d}_{\beta (1,0) a,jm} - \bd{b}^\dagger_{\beta (1,0) a,jm} &=&
\sqrt{2} \sum_{\alpha} M^{~~\alpha}_{\beta}  \bd{D}_{\alpha (1,0)
a,jm} ~~~,
\nonumber \\
\label{eq36} \eeqa
leading to the expressions \beqa {\bd
b}^\dagger_{\beta (1,0) a,jm} & = & \frac{1}{\sqrt{2}} \sum_\alpha
M_\beta^{~\alpha} \left[ {\bd P}^\dagger_{\alpha (1,0) a,jm} - {\bd
D}_{\alpha (1,0) a,jm} \right]
\nonumber \\
{\bd b}_{\beta (0,1) {\bar a},jm} & = &
\frac{1}{\sqrt{2}} \sum_\alpha
M_{~~\beta}^{\alpha}
\left[
{\bd P}_{\alpha (0,1) {\bar a},jm} + {\bd D}^\dagger_{\alpha (0,1) {\bar a},jm}
\right]
\nonumber \\
{\bd d}^\dagger_{\beta (0,1) {\bar a},jm} & = &
\frac{1}{\sqrt{2}} \sum_\alpha
M_{~~\beta}^{\alpha}
\left[
{\bd D}^\dagger_{\alpha (0,1) {\bar a},jm} - {\bd P}_{\alpha (0,1) {\bar a},jm}
\right]
\nonumber \\
{\bd d}_{\beta (1,0) a,jm} & = &
\frac{1}{\sqrt{2}} \sum_\alpha
M_{\beta}^{~\alpha}
\left[
{\bd D}_{\alpha (1,0) a,jm} + {\bd P}^\dagger_{\alpha (1,0) a,jm}
\right]
~~~.
\nonumber \\
\label{eq36a} \eeqa 
A more general transformation can be obtained
introducing a displacement, like in the boson case, 

\beqa
\bd{b}^{\dagger}_{\alpha (1,0) a,jm} \longrightarrow
\sqrt{\lambda_{\alpha}} \bd{b}^{\dagger}_{\alpha (1,0) a,jm}
\nonumber \\
\bd{b}^{\alpha (1,0) a,jm} \longrightarrow \frac{1}{
\sqrt{\lambda_{\alpha}}} \bd{b}^{\alpha (1,0) a,jm} \label{eq37}
\eeqa for the quarks and similarly for the anti-quarks:
\beqa
\bd{d}^{\dagger}_{\alpha (0,1) {\bar a},jm} \longrightarrow
\sqrt{\lambda_{\alpha}} \bd{d}^{\dagger}_{\alpha (0,1) {\bar a},jm}
\nonumber \\
\bd{d}^{\alpha (0,1) {\bar a},jm} \longrightarrow \frac{1}{
\sqrt{\lambda_{\alpha}}} \bd{d}^{\alpha (0,1) {\bar a},jm} ~~~.
\label{eq37a} \eeqa
A further generalization may be defined by
introducing Grassman numbers $\eta_{\alpha (1,0) a,jm}$ and
$\eta_{\alpha (0,1) {\bar a},jm}$ such \beqa {\bd b}^\dagger_{\alpha
(1,0) a,jm} & \rightarrow & \sqrt{\lambda_\alpha} {\bd
b}^\dagger_{\alpha (1,0) a,jm} - \eta_{\alpha (1,0) a,jm}
\nonumber \\
{\bd b}_{\alpha (0,1) {\bar a},jm} & \rightarrow &
\frac{1}{\sqrt{\lambda_\alpha}} {\bd b}_{\alpha (0,1) {\bar a},jm} -
\eta_{\alpha (0,1) {\bar a},jm} \label{eq32bb} \eeqa for the quark
part and similarly for the anti-quark part:
\beqa {\bd
d}^\dagger_{\alpha (0,1) {\bar a},jm} & \rightarrow &
\sqrt{\lambda_\alpha} {\bd d}^\dagger_{\alpha (0,1) {\bar a},jm} -
\eta_{\alpha (0,1) {\bar a},jm}
\nonumber \\
{\bd d}_{\alpha (1,0) a,jm} & \rightarrow &
\frac{1}{\sqrt{\lambda_\alpha}} {\bd d}_{\alpha (1,0) a,jm} -
\eta_{\alpha (1,0) a,jm} \label{eq32bbc} \eeqa The $\eta_{\alpha
(1,0) a,jm}$ ($\eta_{\alpha (0,1) {\bar a},jm}$) are tensors and
transform in the same way as the ${\bd d}$- and ${\bd b}$-operators
with the lower index.

\subsection{Nature of the new fermion vacuum state}
With the redefinitions (\ref{eq37}) and (\ref{eq37a}), we obtain

\beqa 
&  
\sqrt{\lambda_{\alpha}}\bd{d}^{\dagger}_{\alpha (0,1) {\bar
a},jm} -
 \frac{1}{ \sqrt{\lambda_{\alpha}}}\bd{b}_{\alpha (0,1) {\bar a},jm} &
\nonumber \\
 &=&
\nonumber \\
&
 -\sqrt{2}
\sum_{\beta} M^{\beta}_{~\alpha}  \bd{P}_{\beta (0,1) {\bar a},jm}
~~~; &
\nonumber \\
& \sqrt{\lambda_{\alpha}}\bd{b}^{\dagger}_{\alpha (1,0) a,jm} -
 \frac{1}{ \sqrt{\lambda_{\alpha}}}\bd{d}_{\alpha (1,0) a,jm} &
\nonumber \\
 &=&
\nonumber \\
&
 -\sqrt{2}
\sum_{\beta} M^{~~\beta}_{\alpha}  \bd{D}_{\beta (1,0) a,jm} ~~~. &
\label{eq38} \eeqa leading to the equations
\beqa \left(
\sqrt{\lambda_{\alpha}}\bd{d}^{\dagger}_{\alpha (0,1) {\bar a},jm} -
\frac{1}{ \sqrt{\lambda_{\alpha}}} \bd{b}_{\alpha (0,1) {\bar a},jm}
\right) | \tilde{0} \rangle & = & 0
\nonumber \\
\left(  \sqrt{\lambda_{\alpha}}\bd{b}^{\dagger}_{\alpha (1,0) a,jm} -
 \frac{1}{ \sqrt{\lambda_{\alpha}}}\bd{d}_{\alpha (1,0) a,jm} \right)
 \mid {\tilde 0} \rangle & = & 0
~~~, \label{eq39} \eeqa
which imply
\beqa \bd{b}_{\alpha (0,1) {\bar
a},jm}  | \tilde{0} \rangle & = & +\lambda_{\alpha}
\bd{d}^{\dagger}_{\alpha (0,1) {\bar a}} | \tilde{0} \rangle
\nonumber \\
\bd{d}_{\alpha (1,0) a,jm}  | \tilde{0} \rangle & = &
+\lambda_{\alpha} \bd{b}^{\dagger}_{\alpha (1,0) a} | \tilde{0}
\rangle ~~~. 
\label{eq40} 
\eeqa 
The structure of the fermion vacuum
state can be determined in the same way we use for bosons, except
for the restrictions imposed by the Pauli Principle. 
Denoting by $\mid 0\rangle$ the vacuum state of the ${\bd b}$- and
${\bd d}$-operators and by ${\cal N}$ the normalization of the
new vacuum state $\mid {\tilde 0}\rangle$, whose expression is
$\Pi_{\alpha a m} \left(1/\sqrt{1+\lambda_\alpha^2}\right)$, but
not relevant for the following discussion, we have

\beqa 
\mid {\tilde 0} \rangle
& = & {\cal N} \exp{ \left[ \sum_{\alpha}
\lambda_{\alpha}(\bd{d}^{\dagger\alpha} \cdot
\bd{b}^{\dagger}_{\alpha}) \right]}|0 \rangle
\nonumber \\
& = &
{\cal N} \exp{
\left[
\sum_{\alpha am}
\lambda_{\alpha}\bd{d}^{\dagger\alpha (1,0) a,jm}
\bd{b}^{\dagger}_{\alpha (1,0) a,jm}
\right]
}
| 0 \rangle
\nonumber \\
& = &
{\cal N}\Pi_{\alpha am}
\exp{ \left[
\lambda_{\alpha}\bd{d}^{\dagger\alpha (1,0) a,jm}
\bd{b}^{\dagger}_{\alpha (1,0) a,jm}
\right]}
|0 \rangle
\nonumber \\
& = &
{\cal N}\prod_{\alpha am} \left[ 1+(-1)^{\chi_a +j-m} \lambda_{\alpha}
\bd{d}^{\dagger}_{\alpha (0,1)\bar{a},jm}
\bd{b}^{\dagger}_{\alpha (1,0) a,j-m}
\right] |0 \rangle
~~~.
\nonumber \\
\label{eq42} 
\eeqa 
(In the last step we changed $m$ to $-m$.)
In the last line we expanded each exponential up
to products of two terms, taking into account that $\left( {\bd
b}^\dagger_{\alpha (1,0) a,j-m}\right)^2$ = $\left( {\bd
d}^\dagger_{\alpha (0,1) {\bar a},jm}\right)^2=0$. This ansatz
corresponds to a condensate of quark-antiquark-pairs, as expected.
These pairs are coupled to color-spin zero, thus, the trial state
has definite color-spin zero. Now we apply the annihilation operator
${\bd b}_{\alpha (0,1) {\bar a}}$ to this ansatz obtaining

\beqa
\bd{b}_{\alpha (0,1) {\bar a},jm} \mid {\tilde 0} \rangle
&=&
\bd{b}_{\alpha (0,1) {\bar a},jm} {\cal N} \prod_{\alpha ' a'm'}
\left[ 1+(-1)^{\chi_a +j-m'} \lambda_{\alpha '}
\bd{d}^{\dagger}_{\alpha' (0,1) \bar{a}',jm'}
\bd{b}^{\dagger}_{\alpha ' (1,0) a',j-m'}\right] \mid 0 \rangle 
\nn\\
&=&
{\cal N}\left \{ \prod_{(\alpha ', a'm')\ne (\alpha , am)}
\left[ 1+(-1)^{\chi_a +j-m')} \lambda_{\alpha '}
\bd{d}^{\dagger}_{\alpha' (0,1) \bar{a}'jm'}
\bd{b}^{\dagger}_{\alpha ' (1,0) a'j-m'}\right] \right \} \nn\\
&\times&
\bd{b}_{\alpha (0,1) {\bar a},jm}
\left[ 1+(-1)^{\chi_a +j-m} \lambda_{\alpha}
\bd{d}^{\dagger}_{\alpha (0,1) \bar{a},jm}
\bd{b}^{\dagger}_{\alpha (1,0)a,j-m}\right] \mid 0 \rangle \nn\\
&=&
{\cal N}\left \{ \prod_{(\alpha ', a'm')\ne (\alpha , am)}
\left[ 1+(-1)^{\chi_a +j-m'} \lambda_{\alpha '}
\bd{d}^{\dagger}_{\alpha' (0,1) \bar{a}'jm'}
\bd{b}^{\dagger}_{\alpha ' (1,0) a'j-m'}\right] \right \} \nn\\
&\times&
(-1)^{(\chi_a +j-m)+(\chi_a + j +m)} \lambda_{\alpha}
\bd{b}^{\alpha (1,0)a,j-m}
\bd{b}^{\dagger}_{\alpha (1,0)a,j-m}
(-1)\bd{d}^{\dagger}_{\alpha (0,1) {\bar a}jm} \mid 0 \rangle
\nonumber \\
&=&
{\cal N}\left\{
\prod_{(\alpha ', a'm')\ne (\alpha , am)}
\left[ 1+(-1)^{\chi_a +j-m'} \lambda_{\alpha '}
\bd{d}^{\dagger}_{\alpha' (0,1) \bar{a}'jm'}
\bd{b}^{\dagger}_{\alpha ' (1,0)a'j-m'}\right] \right\}
\nonumber \\
&\times&
\lambda_{\alpha}\bd{d}^{\dagger}_{\alpha (0,1) {\bar a}jm} \mid 0 \rangle
~~~.
\label{eq43}
\eeqa
Since a fermion creation-operator commutes with the product of two
fermion creation operators, we have
\beqa
\bd{b}_{\alpha (0,1) {\bar a}jm} \mid {\tilde 0} \rangle
&=&
{\cal N}\left \{ \prod_{(\alpha ', a'm')\ne (\alpha , am)}
\left[ 1+(-1)^{\chi_a +j-m'} \lambda_{\alpha '}
\bd{d}^{\dagger}_{\alpha' (0,1) \bar{a}'m'}
\bd{b}^{\dagger}_{\alpha ' (1,0)a'j-m'}\right] \right \} \nn\\
&\times&
\lambda_{\alpha}\bd{d}^{\dagger}_{\alpha (0,1) {\bar a}jm}
\left[ 1+(-1)^{\chi_a +j-m} \lambda_{\alpha}
{\bd d}^{\dagger}_{\alpha (0,1) \bar{a},jm}
\bd{b}^{\dagger}_{\alpha (1,0) a,j-m}\right] \mid 0 \rangle \nn\\
&=&
\lambda_{\alpha}\bd{d}^{\dagger}_{\alpha (0,1) {\bar a}jm}
\left \{ {\cal N}\prod_{\alpha ' a'm'}
\left[ 1+(-1)^{\chi_a +j-m'} \lambda_{\alpha '}
\bd{d}^{\dagger}_{\alpha' (0,1) \bar{a}'jm'}
\bd{b}^{\dagger}_{\alpha ' (1,0) a'j-m'}\right] \right \}
\mid 0 \rangle \nn\\
&=& \lambda_{\alpha}\bd{d}^{\dagger}_{\alpha (0,1) {\bar a}jm}
\mid {\tilde 0} \rangle
~~~.
\eeqa
We repeat the above steps with the application of the
annihilation operator ${\bd d}$
\beqa
\bd{d}_{\alpha (1,0) a,jm} \mid {\tilde 0} \rangle
&=&
\bd{d}_{\alpha (0,1) a,jm} {\cal N}\prod_{\alpha ' a'm'}
\left[ 1+(-1)^{\chi_a +j+m')} \lambda_{\alpha '}
\bd{d}^{\dagger}_{\alpha' (0,1) \bar{a}',j-m'}
\bd{b}^{\dagger}_{\alpha ' (1,0) a',jm'}\right] \mid 0 \rangle \nn\\
&=&
{\cal N}\left \{ \prod_{(\alpha ', a'm')\ne (\alpha , am)}
\left[ 1+(-1)^{\chi_a +j+m'} \lambda_{\alpha '}
\bd{d}^{\dagger}_{\alpha' (0,1) \bar{a}'j-m'}
\bd{b}^{\dagger}_{\alpha ' (1,0) a'jm'}\right] \right \} \nn\\
&\times&
\bd{d}_{\alpha (1,0) a,jm}
\left[ 1+(-1)^{\chi_a +j+m} \lambda_{\alpha}
\bd{d}^{\dagger}_{\alpha (0,1) \bar{a},j-m}
\bd{b}^{\dagger}_{\alpha (1,0)a,jm}\right] \mid 0 \rangle \nn\\
&=&
{\cal N}\left \{ \prod_{(\alpha ', a'm')\ne (\alpha , am)}
\left[ 1+(-1)^{\chi_a +j+m'} \lambda_{\alpha '}
\bd{d}^{\dagger}_{\alpha' (0,1) \bar{a}'j-m'}
\bd{b}^{\dagger}_{\alpha ' (1,0) a'jm'}\right] \right \} \nn\\
&\times&
(-1)^{2(\chi_a +j+m)} \lambda_{\alpha}
\bd{d}^{\alpha (0,1){\bar a},j-m}
\bd{d}^{\dagger}_{\alpha (0,1)a,j-m}
\bd{b}^{\dagger}_{\alpha (1,0) ajm} \mid 0 \rangle
\nonumber \\
&=&
{\cal N}\left\{
\prod_{(\alpha ', a'm')\ne (\alpha , am)}
\left[ 1+(-1)^{\chi_a +j+m'} \lambda_{\alpha '}
\bd{d}^{\dagger}_{\alpha' (0,1) \bar{a}'j-m'}
\bd{b}^{\dagger}_{\alpha ' (1,0)a'jm'}\right] \right\}
\nonumber \\
&\times&
(+1)\lambda_{\alpha}\bd{b}^{\dagger}_{\alpha (1,0) ajm} \mid 0 \rangle
~~~.
\label{eq43x}
\eeqa
Using again that a fermion creation operator commutes with the
product of two fermion creation operators, we have

\beqa
\bd{d}_{\alpha (1,0) a,jm} \mid {\tilde 0} \rangle
&=&
{\cal N}\left \{ \prod_{(\alpha ', a'm')\ne (\alpha , am)}
\left[ 1+(-1)^{\chi_a +j+m'} \lambda_{\alpha '}
\bd{d}^{\dagger}_{\alpha' (0,1) \bar{a}'-m'}
\bd{b}^{\dagger}_{\alpha ' (1,0)a'jm'}\right] \right \} \nn\\
&\times&
\lambda_{\alpha}\bd{b}^{\dagger}_{\alpha (1,0) a,jm}
\left[ 1+(-1)^{\chi_a +j+m} \lambda_{\alpha}
{\bd d}^{\dagger}_{\alpha (0,1) \bar{a},j-m}
\bd{b}^{\dagger}_{\alpha (1,0) a,jm}\right] \mid 0 \rangle \nn\\
&=&
+\lambda_{\alpha}\bd{b}^{\dagger}_{\alpha (1,0) a,jm}
\left \{ {\cal N}\prod_{\alpha ' a'm'}
\left[ 1+(-1)^{\chi_a +j+m'} \lambda_{\alpha '}
\bd{d}^{\dagger}_{\alpha' (0,1) \bar{a}'jm'}
\bd{b}^{\dagger}_{\alpha ' (1,0) a'j-m'}\right] \right \}
\mid 0 \rangle\nn\\
&=& +\lambda_{\alpha}\bd{b}^{\dagger}_{\alpha (1,0) ajm}\mid {\tilde 0} \rangle
~~~.
\eeqa
This proves that our ansatz satisfies the operator equation for the
new fermionic vacuum. The vacuum has definite color and spin, when
we assume that the indices of the transformation  matrix do not
depend on color nor on the spin quantum numbers. With this, the
${\bd P}$ and ${\bd D}$-operators are given by
\beqa
{\bd P}^\dagger_{\alpha (1,0) a,jm} & = & \frac{1}{\sqrt{2}}\sum_\beta
M_\alpha^{~\beta}
\left( \sqrt{\lambda_\beta}{\bd b}^\dagger_{\beta (1,0) a,jm} +
\frac{1}{\sqrt{\lambda_\beta}}{\bd d}_{\beta (1,0) a,jm} \right)
\nonumber \\
{\bd P}_{\alpha (0,1) {\bar a},jm} & = &
\frac{1}{\sqrt{2}}\sum_\beta M_{~~\alpha}^\beta
\left( \frac{1}{\sqrt{\lambda_\beta}}{\bd b}_{\beta (0,1) {\bar a},jm} -
\sqrt{\lambda_\beta}{\bd d}^\dagger_{\beta (0,1) {\bar a},jm}
\right)
\nonumber \\
{\bd D}^\dagger_{\alpha (0,1) {\bar a},jm} & = &
\frac{1}{\sqrt{2}}\sum_\beta
M_{~~\alpha}^{\beta}
\left( \sqrt{\lambda_\beta}{\bd d}^\dagger_{\beta (0,1) {\bar a},jm} +
\frac{1}{\sqrt{\lambda_\beta}}{\bd b}_{\beta (0,1) {\bar a},jm} \right)
\nonumber \\
{\bd D}_{\alpha (1,0) a,jm} & = &
\frac{1}{\sqrt{2}}\sum_\beta M_{\alpha}^{~\beta}
\left( \frac{1}{\sqrt{\lambda_\beta}}{\bd d}_{\beta (1,0) a,jm} -
\sqrt{\lambda_\beta}{\bd b}^\dagger_{\beta (1,0) a,jm}
\right)
~~~.
\label{eq32a}
\eeqa
Adding a displacement (by adding a Grassman variable), the equations
for the ${\bd P}$-operators change to
\beqa
{\bd P}^\dagger_{\alpha (1,0) a,jm} & = &
\frac{1}{\sqrt{2}}\sum_\beta M_\alpha^{~\beta}
\left( \sqrt{\lambda_\beta}{\bd b}^\dagger_{\beta (1,0) a,jm} +
\frac{1}{\sqrt{\lambda_\beta}}{\bd d}_{\beta (1,0) a,jm} \right)
\nonumber \\
&& -\sqrt{2}\sum_\beta M_\alpha^{~\beta}\eta_{\beta (1,0) a,jm}
\nonumber \\
{\bd P}_{\alpha (0,1) {\bar a}jm} & = & \frac{1}{\sqrt{2}}\sum_\beta
M_{~~\alpha}^{\beta}
\left( \frac{1}{\sqrt{\lambda_\beta}}
{\bd b}_{\beta (0,1) {\bar a},jm} -
\sqrt{\lambda_\beta}
{\bd d}^\dagger_{\beta (0,1) {\bar a},jm} \right)
\nonumber \\
{\bd D}^\dagger_{\alpha (0,1) {\bar a},jm} & = &
\frac{1}{\sqrt{2}}\sum_\beta M_{~~\alpha}^{\beta}
\left( \sqrt{\lambda_\beta}{\bd d}^\dagger_{\beta (0,1) {\bar a},jm} +
\frac{1}{\sqrt{\lambda_\beta}}{\bd b}_{\beta (0,1) {\bar a},jm} \right)
\nonumber \\
&& -\sqrt{2}\sum_\beta M^\beta_{~\alpha} \eta_{\beta (0,1) {\bar a},jm}
\nonumber \\
{\bd D}_{\alpha (1,0) ajm} & = & \frac{1}{\sqrt{2}}\sum_\beta
M_{\alpha}^{~\beta}
\left( \frac{1}{\sqrt{\lambda_\beta}}
{\bd d}_{\beta (1,0) a,jm} -
\sqrt{\lambda_\beta}
{\bd b}^\dagger_{\beta (1,0) a,jm} \right)
~~~.
\label{eq32aax}
\eeqa
In total, we have now the following parameters to vary:
\beqa &
M_\alpha^{~\beta} ~,~ \lambda_\alpha ~{\rm and}~ \eta_{\alpha (1,0)
a,jm} ~,~ \eta_{\alpha (0,1) {\bar a},jm} ~~~. \label{eq32dd} \eeqa
Finally, we give the expressions of the former creation and
annihilation operators in terms of the new ones, including the
information on $\lambda_\alpha$, $\eta_{\alpha (1,0) a, jm}$ and
$\eta_{\alpha (0,1) {\bar a}, jm}$:

\beqa
{\bd b}^\dagger_{\beta
(1,0) a,jm} & = & \frac{1}{\sqrt{2\lambda_\beta}} \sum_\alpha
M_\beta^{~\alpha} \left[ {\bd P}^\dagger_{\alpha (1,0) a,jm} - {\bd
D}_{\alpha (1,0) a,jm} \right]
\nonumber \\
&& + \frac{\eta_{\beta (1,0) a,jm}}{\sqrt{\lambda_\beta}}
\nonumber \\
{\bd b}_{\beta (0,1) {\bar a},jm} & = &
\sqrt{\frac{\lambda_\beta}{2}} \sum_\alpha
M_{~~\beta}^{\alpha}
\left[
{\bd P}_{\alpha (0,1) {\bar a},jm} + {\bd D}^\dagger_{\alpha (0,1) {\bar a},jm}
\right]
\nonumber \\
&& + \sqrt{\lambda_\beta}\eta_{\beta (0,1) {\bar a},jm}
\nonumber \\
{\bd d}^\dagger_{\beta (0,1) {\bar a},jm} & = &
\frac{1}{\sqrt{2\lambda_\beta}} \sum_\alpha
M_{~~\beta}^{\alpha}
\left[
{\bd D}^\dagger_{\alpha (0,1) {\bar a},jm} - {\bd P}_{\alpha (0,1) {\bar a},jm}
\right]
\nonumber \\
&& + \frac{\eta_{\beta (0,1) {\bar a},jm}}{\sqrt{\lambda_\beta}}
\nonumber \\
{\bd d}_{\beta (1,0) a,jm} & = &
\sqrt{\frac{\lambda_\beta}{2}} \sum_\alpha
M_{\beta}^{~\alpha}
\left[
{\bd D}_{\alpha (1,0) a,jm} + {\bd P}^\dagger_{\alpha (1,0) a,jm}
\right]
\nonumber \\
&& + \sqrt{\lambda_\beta}\eta_{\beta (1,0) a,jm}
~~~.
\label{eq36xy}
\eeqa
\section{Relation to the coherent state as used in \cite{adam-trial-1,adam-trial-2}}
In \cite{adam-trial-1,adam-trial-2} a monopole gas for the
description of the QCD ground state was proposed and thermodynamical
properties extracted. The trial state of this monopole gas,
depending on several parameters, has the form of a coherent state.
In this section we show that this trial state can be recast in the
many-body language of the previous sections. The trial state
introduced in \cite{adam-trial-1,adam-trial-2} has the form

\beqa
\langle A \mid {\tilde 0} \rangle & = & {\cal N}
e^{-\frac{1}{2}\int d^3x \int d^3 y A(x)\omega (x-y) A(y)} ~~~,
\label{trial-adam}
\eeqa
where ${\cal N}$ is a normalization. We
start with the expression of the field components $A(x)_a$ in the
basis of the operators $\bd b^\dagger$ and $\bd b$
\beqa
A(x)_a & =
& \sum_\alpha \phi^\alpha  (x) \frac{1}{\sqrt{2}}\left( {\bd
b}^\dagger_{\alpha a} + {\bd b}_{\alpha a} \right)
\nonumber \\
A^\dagger(x)_a~=~A^a (x) & = & \sum_\alpha \phi_\alpha  (x)
\frac{1}{\sqrt{2}}\left( {\bd b}^{\dagger\alpha a} + {\bd b}^{\alpha
a} \right) ~~~,
\label{eq4}
\eeqa
The amplitude $\phi^\alpha (x)$
is to be associated with the factor
 $\xi_{{\bar \alpha}}$ of  \cite{adam-trial-1}, and

\beqa
\phi^*_\alpha (x) & = & \phi^\alpha (x) ~=~(-1)^{\varphi
(\alpha)} \phi_{{\bar \alpha}} ~~~.
\label{4a}
\eeqa
In the
quantization of the field $A_a(x)$ there appears an additional
factor $1/\sqrt{\Omega_\alpha}$, with $\Omega_\alpha$ being the
frequency of the solution $\phi^\alpha (x)$. We adopt the
notation that this factor is included in the function. In order to
preserve gauge invariance \cite{kogan}, $\phi^\alpha (x)$ has to
contain a combination of perturbative $A$-fields with a monopole
solution, i.e., in general, different boson operators ${\bd
b}^\dagger_{\alpha a}$ have to be defined. For simplicity, we use
one generic expression.

With this, we have

\beqa
& A^\dagger_a(x) \omega (x-y) A_a(y)  = &
\nonumber \\
& \frac{1}{\sqrt{2}}
\sum_a\sum_{\alpha} \phi_\alpha (x) \left( {\bd b}^{\dagger\alpha a}
+ {\bd b}^{\alpha a} \right) \omega (x-y) &
\nonumber \\
&
\sum_{\beta} \phi^\beta (y) \left( {\bd b}^\dagger_{\beta a}
+ {\bd b}_{\beta a} \right) \omega (x-y) &
\nonumber \\
& = &
\nonumber \\
& \frac{1}{\sqrt{2}}
\sum_a\sum_{\alpha \beta}
\phi_\alpha (x) \omega(x-y) \phi^\beta (y)
&
\nonumber \\
& \left( {\bd b}^{\alpha a}{\bd b}^\dagger_{\beta a} +{\bd
b}^{\dagger \alpha a}{\bd b}_{\beta a} +{\bd b}^{\dagger \alpha
a}{\bd b}^\dagger_{\beta a} +{\bd b}^{\alpha a}{\bd b}_{\beta a}
\right) ~~~.&
\label{eq5}
\eeqa
where, in lowering indexes and
contracting them we have used the relation

\beqa
\phi_\alpha
(x){\bd b}^{\alpha a} & = & \phi_\alpha (x) (-1)^{\varphi (\alpha
)} {\bd b}^{a}_{{\bar \alpha}}
\nonumber \\
& = & \phi^{{\bar \alpha}} (x) {\bd b}^{a}_{{\bar \alpha}} ~~~,
\label{eq6} \eeqa

With this, (\ref{eq5}) can be rewritten as

\beqa
& \sum_a
A^\dagger_a(x) \omega (x-y) A_a(y)  = &
\nonumber \\
& \frac{1}{\sqrt{2}}
\sum_a\sum_{\alpha \beta}
\phi^{{\bar \alpha}} (x) \omega(x-y) \phi^\beta (y)
&
\nonumber \\
& \left( {\bd b}_{{\bar \alpha}}^{a}{\bd b}^\dagger_{\beta a} +{\bd
b}^{\dagger a}_{{\bar \alpha}} {\bd b}_{\beta a} +{\bd b}^{\dagger
a}_{{\bar \alpha}} {\bd b}^\dagger_{\beta a} +{\bd b}^a_{{\bar
\alpha}}{\bd b}_{\beta a} \right) ~~~.&
\label{eq8}
\eeqa
In terms
of the scalar products,

\beqa
\sum_a {\bd b}^a_{{\bar \alpha}} {\bd
b}^{\dagger}_{\beta a} & = &  \left( {\bd b}_{{\bar \alpha}} \cdot
{\bd b}^{\dagger}_{\beta} \right) ~~~.
\label{eq8a}
\eeqa
it reads
\beqa
& \sum_a A^\dagger_a(x) \omega (x-y) A_a(y) = &
\nonumber \\
& \frac{1}{\sqrt{2}}
\sum_{\alpha \beta}
\phi^{{\bar \alpha}} (x) \omega(x-y) \phi^\beta (y)
&
\nonumber \\
& \left\{ \left( {\bd b}_{{\bar \alpha}} \cdot {\bd
b}^\dagger_{\beta } \right) +\left( {\bd b}^{\dagger}_{{\bar
\alpha}} \cdot {\bd b}_{\beta} \right) +\left( {\bd
b}^{\dagger}_{{\bar \alpha}} \cdot {\bd b}^\dagger_{\beta} \right)
+\left( {\bd b}_{{\bar \alpha}} \cdot {\bd b}_{\beta} \right)
\right\} ~~~. & 
\label{eq8b} 
\eeqa 
Because $\alpha$ is a dummy
index, we can skip the bar over $\alpha$ in the sum.

By integrating Eq. (\ref{eq5}) over $x$ and $y$ we obtain the same
structure as in (\ref{eq8b}), by writing

\beqa
A^{\alpha \beta} & =
& \int dx dy \phi^\alpha (x) \omega (x-y) \phi^\beta (y) ~~~.
\label{eq10}
\eeqa
for the factors in front of the scalar products.
Using the ansatz for monopoles given in
\cite{adam-trial-1,adam-trial-2}, the factors $A^{\alpha \beta}$ can
be determined. What we have shown is that the coherent trial state,
as used in \cite{adam-trial-1,adam-trial-2}, can be cast into a
standard language, like in conventional many body theories.

\subsection{Taking the product of the operators ${\bd P}^\dagger$ and ${\bd
P}$} 

The expression (\ref{eq8}) for the exponent can finally be
rewritten as a product of creation and annihilation operators by
taking the product of two ${\bd P}$-operators as 
\beqa 
&
\sum_a\sum_{\alpha} \lambda_\alpha {\bd P}^{\dagger}_{\alpha a} {\bd
P}^{\alpha a} = &
\nonumber \\
& \frac{1}{2}
\sum_a\sum_{\alpha} \lambda_\alpha \sum_{\beta_1 \beta_2}
\left( M_{\alpha}^{~\beta_1} {\bd b}^{\dagger}_{\beta_{1}a}
+ N_{\alpha}^{~\beta_1} {\bd b}_{\beta_1 a}  \right)
&
\nonumber \\
& \left( M^{\alpha}_{~\beta_2} {\bd b}^{\beta_2 a} -
N^{\alpha}_{~\beta_2} {\bd b}^{\dagger\beta_2 a}  \right) ~~~.
\label{eq11} \eeqa 
Performing the explicit multiplication gives
\beqa 
& 
\sum_{\beta_1 \beta_2} \left\{ A^{\beta_1}_{\beta_2}
\left( {\bd b}^{\dagger}_{\beta_1} \cdot {\bd b}^{\beta_2}\right) -
B^{\beta_1}_{\beta_2} \left({\bd b}^{\dagger}_{\beta_1} \cdot {\bd
b}^{\dagger\beta_2}\right) \right. &
\nonumber \\
& \left. + B^{\prime\beta_1}_{\beta_2} \left({\bd b}_{\beta_1} \cdot
{\bd b}^{\beta_2} \right) 
- A^{\prime\beta_1}_{\beta_2} \left( 
{\bd b}_{\beta_1} \cdot {\bd b}^{\dagger\beta_2}\right) \right\} & ~~~,
\label{eq13} 
\eeqa 
with

\beqa
A^{\beta_1}_{\beta_2} & = & \frac{1}{2}\sum_\alpha \lambda_\alpha
M_{\alpha}^{~\beta_1} M^{\alpha}_{~\beta_2}
\nonumber \\
B^{\beta_1}_{\beta_2} & = & \frac{1}{2}\sum_\alpha \lambda_\alpha
M_{\alpha}^{~\beta_1} N^{\alpha}_{~\beta_2}
\nonumber \\
B^{\prime\beta_1}_{\beta_2} & = & \frac{1}{2}
\sum_\alpha \lambda_\alpha
N_{\alpha}^{~\beta_1} M^{\alpha}_{~\beta_2}
\nonumber \\
A^{\prime\beta_1}_{\beta_2} & = & \frac{1}{2}
\sum_\alpha \lambda_\alpha
N_{\alpha}^{~\beta_1} N^{\alpha}_{~\beta_2}
~~~.
\label{eq14}
\eeqa

In order to get the same structure as in (\ref{eq8b}) -
(\ref{eq10}), the upper indices in the ${\bd b}$ operators in
(\ref{eq14}) are lowered and the corresponding lower indices in the
coefficient matrices are raised. In addition, the canonical transformation 
${\bd b}^\dagger_{\alpha a}$ $\rightarrow$ $i{\bd b}^\dagger_{\alpha a}$,
${\bd b}^{\alpha a}$ $\rightarrow$ $\frac{1}{i}{\bd b}^{\alpha a}$ is
performed, which maintains the commutation relations. This transformation changes the 
sign in front of the products of two creation or two annihilation operators, but
not in the product of a creation with an annihilation operator. 
With this, we have

\beqa 
& 
\sum_{\beta_1 \beta_2} \left\{ A^{\beta_1\beta_2} 
\left( {\bd b}^{\dagger}_{\beta_1} \cdot {\bd b}_{\beta_2}\right) +
B^{\beta_1\beta_2} \left({\bd b}^{\dagger}_{\beta_1} \cdot {\bd
b}^{\dagger}_{\beta_2}\right) \right. &
\nonumber \\
& \left. 
+ B^{\prime\beta_1\beta_2} \left({\bd b}_{\beta_1} \cdot
{\bd b}_{\beta_2}\right) + A^{\prime\beta_1\beta_2} 
\left( {\bd b}_{\beta_1} \cdot {\bd b}^{\dagger}_{\beta_2}\right) \right\} &
~~~, 
\label{eq14a} 
\eeqa

We choose the particular relations

\beqa
A^{\beta_1\beta_2} & = & B^{\beta_1\beta_2}~=~B^{\prime\beta_1\beta_2}
~=~A^{\prime\beta_1\beta_2}
~~~,
\label{eq15}
\eeqa
thus, leading to (\ref{eq8b}) with the integration (\ref{eq10}) performed. 
(\ref{eq15}) demonstrates that the ansatz of a monopole gas is contained in
the many-body trial state, but eliminating the conditions
(\ref{eq15}) permits a more general structure, which is of great advantage.

The QCD trial state can now be formally written as 
\beqa
e^{\sum_{\alpha} \lambda_\alpha \left( {\bd P}^\dagger_\alpha \cdot {\bd P}^\alpha
\right)} \mid {\tilde 0} \rangle  ~~~. 
\label{trial-fin} 
\eeqa
Because ${\bd P}^{\alpha a}\mid {\tilde 0} \rangle = 0$, only the
first term in the exponential contributes. Thus, the trial state
{\it has been reduced to the vacuum state} $\mid {\tilde 0}
\rangle$,

The matrix elements $M_\alpha^{~\beta}$ (= $N_\alpha^{~\beta}$) can be dealt with in
two ways:\\
i) Simply as parameters which are determined by minimizing the
expectation value of the Hamiltonian with respect to
a trial state, or \\
ii) determine $M_\alpha^{~\beta}$ via the $A^{\alpha\beta}$ of Eq.
(\ref{eq10}) using also the relations (\ref{eq14})-(\ref{eq15}),
i.e., the matrix elements depend then on the parameters of the
monopole gas. The above relation may be affected by spin mixing,
because when the indices $\alpha$ and $\beta$ contain the total spin
$j$, the matrices $M_\alpha^\beta$ will mix the spin and the trial
state does not have definite spin, though the new vacuum may have
definite spin. If we want to avoid this, $\alpha$ and $\beta$ should
not depend on the total spin.

\section{Conclusions}

In this contribution we have developed a variational procedure which
includes simultaneously fermions and bosons. We use explicitly the
case of non-perturbative QCD \cite{adam-trial-1,adam-trial-2}, as a
working example,  but it can be generalized to any system of
fermions and bosons. A series of trial states have been proposed,
with increasing complexity. They include a simple unitary
transformation plus a possible re-scaling of the boson and fermion
operators and a shift of the boson operators.

We have shown that the trial state presented in
\cite{adam-trial-1,adam-trial-2}, where a coherent state written in
terms of a QCD functional was used in order to describe the ground
state of QCD as a monopole gas, can be recast into a standard
many-body framework. We think that this connection may facilitate
the use of such techniques in dealing with the low energy domain of
QCD. Further work is in progress concerning the use of effective
QCD-inspired Hamiltonians.

\section*{Acknowledgements}

We gratefully acknowledge financial help from DGAPA-PAPIIT (no.
IN103212), from the National Research Council of Mexico (CONACyT)
and DGAPA. P.O.H. thanks the FIAS and the GSI for the hospitality
and the excellent working atmosphere during his sabbatical stay in
Germany. This work has been partially supported by the CONICET and
ANPCyT of Argentina.

\section*{Appendix}

In this appendix we derive properties of the fermion creation
operators under lowering and raising their indices. We start from
the convention we have introduced in \cite{paper-2010}, which is

\beqa {\bd b}^{\dagger -\xi \alpha (0,1) {\bar a}, jm} & = &
(-1)^{\frac{1}{2}+\xi} (-1)^{\chi_a + j-m} {\bd b}^\dagger_{+\xi
\alpha (1,0) a, j-m} ~~~.
\nonumber \\
\label{a1} \eeqa
($\alpha$, $\beta$ correspond to Cartesian indices). In the above
equation (\ref{a1}), $a$ is the color index, $j$ is the spin, $m$
its projection and $\xi$ is the pseudo-spin component. The
pseudo-spin quantum number refers to the upper and lower level and
for the upper level $\xi = \frac{1}{2}$ and for the lower level it
is $\xi = -\frac{1}{2}$. The phase factor $\chi_a$ is always
integer, while $j$ and $m$ are half integer.

The property of the annihilation operator, under raising and
lowering indices, is obtained by the hermitian conjugation of
(\ref{a1})

\beqa \left( {\bd b}^{\dagger -\xi \alpha (0,1) {\bar a}, jm}
\right)^\dagger & = & {\bd b}_{-\xi \alpha (0,1) {\bar a},jm}
\nonumber \\
&& \left[ (-1)^{\frac{1}{2}+\xi+\chi_a + j-m} {\bd b}^\dagger_{\xi
\alpha (1,0) a, j-m} \right]^\dagger
\nonumber \\
& = & (-1)^{\frac{1}{2}+\xi+\chi_a + j-m} {\bd b}^{\xi \alpha (1,0)
a, j-m} ~~~.
\nonumber \\
\label{a1a} \eeqa

Identifying the fermion creation operator with component
$\pm\frac{1}{2}$ with the creation of a particle (${\bd b}^\dagger$)
and the annihilation of an anti-particle ( ${\bd d}$), respectively,
and similarly for the fermion annihilation operator , we arrive at
the following expressions

\beqa {\bd b}^{\dagger \alpha (0,1) {\bar a}, jm} & = & (-1)^{\chi_a
+j+m} {\bd b}^\dagger_{\alpha (1,0) a, j -m}
\nonumber \\
{\bd b}^{\dagger}_{\alpha (1,0) a, jm} & = & (-1)^{\chi_a +j-m} {\bd
b}^{\dagger \alpha (0,1) {\bar a}, j -m}
\nonumber \\
{\bd b}^{\alpha (1,0) a, jm} & = & (-1)^{\chi_a +j-m} {\bd
b}_{\alpha (0,1) {\bar a}, j -m}
\nonumber \\
{\bd b}_{\alpha (0,1) {\bar a}, jm} & = & (-1)^{\chi_a +j+m} {\bd
b}^{\alpha (1,0) a, j -m} ~~~.
\nonumber \\
\label{a3} \eeqa

A similar property holds for the anti-particle operators, i.e.,

\beqa {\bd d}^{\alpha (0,1) {\bar a}, jm} & = & (-1)^{\chi_a +j-m}
{\bd d}_{\alpha (1,0) a, j -m}
\nonumber \\
{\bd d}_{\alpha (1,0) a, jm} & = & (-1)^{\chi_a +j+m} {\bd
d}^{\alpha (0,1) {\bar a}, j -m}
\nonumber \\
{\bd d}^{\dagger \alpha (1,0) a, jm} & = & (-1)^{\chi_a +j+m} {\bd
d}^{\dagger}_{\alpha (0,1) {\bar a}, j -m}
\nonumber \\
{\bd d}^{\dagger}_{\alpha (0,1) {\bar a}, jm} & = & (-1)^{\chi_a
+j-m} {\bd d}^{\dagger \alpha (1,0) a, j -m}
\nonumber \\
~~~. \label{a4} \eeqa Here, one has to include an additional change
in sign due to the phase $(-1)^{\frac{1}{2} - \xi}$.

Note, that the resulting phase property is {\it the same} for the
particle creation as for the anti-particle creation operator. The
same holds for the annihilation operators.

With the help of the just obtained results, we can study the
structure of the ${\bd D}$- and ${\bd D}^\dagger$-operators (see
Section III). The creation operator ${\bd D}^\dagger$ corresponds to
a particle-annihilation operator at negative energy. Thus we make a
similar ansatz for it, as for the annihilation operator ${\bd P}$,
but with a different sign, i.e., \beqa {\bd D}^{\dagger \alpha (1,0)
a,jm} & = & \frac{1}{\sqrt{2}} \sum_\beta M_{\beta}^{~\alpha} \left(
{\bd d}^{\dagger \beta (1,0) a,jm} - {\bd b}^{\beta (1,0) a,jm}
\right) ~~~.
\nonumber \\
\label{a5} \eeqa It can be easily shown that this anti-commutes with
${\bd P}^\dagger$. Now, we lower the index in (\ref{a5}), leading to
\beqa
{\bd D}^{\dagger \alpha (1,0) a,jm} & = & \frac{1}{\sqrt{2}}
\sum_\beta M_\beta^{~\alpha} \left[ (-1)^{\chi_a+j+m} {\bd
d}^\dagger_{\beta (0,1) {\bar a}, j-m} - (-1)^{\chi_a+j-m} {\bd
b}_{\beta (0,1) {\bar a}, j-m} \right]
\nonumber \\
& = & (-1)^{\chi_a+j+m}  \frac{1}{\sqrt{2}} \sum_\beta
M_\beta^{~\alpha} \left( {\bd d}^\dagger_{\beta (0,1) {\bar a}, j-m}
+ {\bd b}_{\beta (0,1) {\bar a}, j-m} \right)
\nonumber \\
& = & (-1)^{\chi_a+j+m} {\bd D}^\dagger_{\alpha (0,1) a, j-m} ~~~.
\nonumber \\
\label{a6} \eeqa
The matrices $M$ satisfy $M_\alpha^{~\beta} = M^\alpha_{~\beta}$.
From this we conclude that the correct ansatz for the new
anti-particle operator is

\beqa {\bd D}^{\dagger}_{\alpha (0,1) a,jm} & = & \frac{1}{\sqrt{2}}
\sum_\beta M^\beta_{~\alpha} \left( {\bd d}^\dagger_{\beta (0,1)
{\bar a},jm} + {\bd b}_{\beta (0,1) {\bar a},jm} \right) ~~~.
\nonumber \\
\label{a7} \eeqa A similar manipulation for the anti-particle
annihilation operator \beqa {\bd D}_{\alpha (1,0) a,jm} & = &
\frac{1}{\sqrt{2}} \sum_\beta M_\alpha^{~\beta} \left( {\bd
d}_{\beta (1,0) a,jm} - {\bd b}^\dagger_{\beta (1,0) a,jm} \right)
\nonumber \\
\label{a8} \eeqa yields \beqa {\bd D}^{\alpha (0,1) {\bar a},jm} & =
& \frac{1}{\sqrt{2}} \sum_\beta M^{\alpha}_{~\beta} \left( {\bd
d}^{\beta (0,1) {\bar a},jm} + {\bd b}^{\dagger \beta (0,1) {\bar
a},jm} \right)
\nonumber \\
\label{a9}
\eeqa
and

\beqa
{\bd D}_{\alpha (1,0) a, jm} & = & (-1)^{\chi_a + j + m} {\bd
D}^{\alpha (0,1) {\bar a},j-m} ~~~.
\label{a10}
\eeqa

Equivalent considerations are applied for the raising and lowering
of the indices of the ${\bd P}^\dagger$ and ${\bd P}$ operators,
with the result:

\beqa
{\bd P}^\dagger_{\alpha (1,0)a,jm} & = & (-1)^{\chi_a+j-m} {\bd P}^{\dagger \alpha (0,1) {\bar a},j-m}
\nonumber \\
{\bd P}_{\alpha (0,1){\bar a},jm} & = & (-1)^{\chi_a+j+m} {\bd P}^{\alpha (1,0) a,j-m}
\label{a11}
\eeqa

\end{document}